\documentclass[aps,pra,twocolumn,showpacs,floatfix]{revtex4-1}
\usepackage{hyperref}

\usepackage{graphicx}
\usepackage[english]{babel}
\usepackage{amsmath}
\usepackage{amssymb}

\newcommand{\mean}[1]{\langle #1 \rangle}

\def\eps{\epsilon}

\def\q{{\bf q}}

\def\w{\omega}

\def\calA{{\cal A}}
\def\calD{{\cal D}}
\def\calF{{\cal F}} 
\def\calG{{\cal G}}

\def\calL{{\cal L}} 
\def\calM{{\cal M}}

\def\bw{\bar \w}
\def\bg{\bar \gamma}
\def\bE{\bar E}
\def\bgn{\bar{g}\bar{n}_0}
\def\G{\Gamma}
\def\bk{{\bf k}}

\begin{document}
\title{Quench Dynamics in Bose condensates in the Presence of a Bath:
Theory and Experiment} 

\author{Adam Ran\c{c}on} 
\author{Chen-Lung Hung}
\altaffiliation[current address : ]{Norman Bridge Laboratory of Physics 12-33, California Institute of Technology, Pasadena, California 91125, USA}
\author{ Cheng Chin}
\author{K. Levin}

\affiliation{James Franck Institute and Department of Physics,
University of Chicago, Chicago, Illinois 60637, USA
}

\begin{abstract} 
In this paper we study the transient dynamics of a Bose superfluid subsequent to
an interaction quench. Essential for equilibration is a source of dissipation which
we include following the approach of Caldeira and Leggett.
Here we solve the equations of motion exactly by integrating out an 
environmental bath. We thereby derive precisely
the time dependent density correlation functions
with the appropriate analytic and asymptotic properties. The resulting structure factor exhibits
the expected damping and thereby differs 
from that of strict Bogoliubov theory.
These damped sound modes, which reflect the
physics beyond mean field approaches, are
characterized and the structure factors are
found to compare favorably with
experiment.
\end{abstract}

\maketitle

Understanding out-of-equilibrium dynamics and dissipation in superfluids
has experienced a revival with recent experiments in cold atoms. These atomic systems afford
access
to new probes of non-equilibrium behavior, not available in condensed matter counterparts,
among these a sudden change of the
interaction strength
\cite{Donley2001,Greiner2005}.
The equilibration process after this ``quench" necessarily involves a source of dissipation,
the understanding of which can elucidate essential microscopic processes, not included
in standard mean field approaches to superfluidity.
The importance of understanding quantum coherence and dissipation was emphasized in
the seminal work of Leggett 
and Caldeira 
\cite{Caldeira1983} and subsequently explored by many others \cite{Ford1987,Grabert1988,Bonart2013,Bonart2012,Chang1985}. 
The goal of this paper is to apply these important ideas to
the quench dynamics of a Bose superfluid and to show how to address related experiments \cite{Hung2012}. 
Following earlier work \cite{Chang1985,Tan2004} we develop a formalism for
 calculations of the real time dynamics of a superfluid 
coupled to a rather general bath.

This formalism is then applied
to address experimental studies of two dimensional Bose gases \cite{Hung2012}. In these experiments, one has direct access to the equal time density correlation functions, or structure factor $S_\mathbf{k}(t)$.  A key experimental observation was that the quench appears to excite acoustic waves, which interfere in both the spatial and temporal domains, leading to Sakharov oscillations. A simple Bogoliubov-level theory was applied to analyze these experiments (see also \cite{Natu2012}), while leaving a few experimental features, such as damping in the Sakharov oscillations, unexplained. 
Here, in investigating the physics of dissipation, we
re-enforce the observation of oscillatory behavior by incorporating the presence of 
damping in
the data analysis. Our study supports the earlier
observation of oscillatory sound modes for some range of $\mathbf{k}$ and $t$ in $S_\bk(t)$.  It also provides insights into why the simplest Bogoliubov-based scheme is more inadequate in situations in which the coupling constant $g$ is suddenly increased. Moreover, our tractable formalism for treating such dissipation, should be a first step in developing tools for elucidating  the microscopics of cold gases which go beyond the simplest mean field theories of the steady state.

The subject and origin of quantum dissipation has a long history. In the Leggett Caldeira (LC) approach
the bath is modeled by an infinite set of harmonic oscillators;  the Hamiltonian is then quadratic
and one can solve the equations of motion exactly by effectively integrating out
the environment.
One similarly introduces a source of ``noise" or dissipation into 
fermionic superfluids via time dependent Ginsburg-Landau theory \cite{Ullah1990}. The counterpart in
bosonic superfluids has been addressed in the context of
stochastic versions of
the  Gross-Pitaevski equation (SGPE) \cite{Blakie2008,Stoof1997,Prokakis}, where more microscopic
approaches to the noise source have been highlighted.

In all these superfluids, in the simplest terms the noise or dissipation corresponds to fluctuations
or processes not included in mean field theory. 
Specifically,
for Bose superfluids, the mechanism for irreversible loss of energy can be
associated with the interaction between 
Bogoliubov quasi-particles. There are microscopic schemes to address these
beyond-Bogoliubov damping effects due to Beliaev \cite{Beliaev1958,Beliaev1958a}. Their
inclusion into dynamics 
is usually via the Schwinger-Keldysh formalism
which can be rather involved with the condensate wave function appearing
non-linearly in functionals that are non-local in space and time.
In general, extensive numerical simulations are necessary, making such schemes
less physically transparent and often restricting their use to rather high temperatures \cite{Blakie2008,Stoof1997}.

The alternative more macroscopic
concept of the environmental bath, applied here, is to split the total system into
two parts: the quantum system where dissipation occurs (say the Bose superfluid
at the Bogoliubov level) and a so-called
environment. Evidence for universality suggests \cite{Chang1985}
that the particular description of
the bath will not affect the essential features of the
dissipative process.
The latter is often modeled presuming Ohmic dissipation.

%

Throughout this paper
we neglect trap effects; our focus is on reasonably short times
where the trap geometry is not important.
In the absence of the bath (as well as trap), the Hamiltonian of the Bogoliubov modes is given by 
$\hat H_{{\rm bog}}(g) =$
\begin{equation}
\sum_\bk \big[\hat \psi^\dag_\bk (\eps_\bk-\mu+2gn_0) \hat \psi_\bk + \frac{gn_0}{2} \hat \psi_\bk\hat \psi_{-\bk} \nonumber \\
+ \frac{gn_0}{2} \hat \psi^\dag_\bk\hat \psi^\dag_{-\bk}\big],
\end{equation}
where $\hat \psi^{(\dag)}_\bk$ annihilates (creates) an atom with momentum $\bk\neq0$ (the dispersion $\eps_\bk$ can be quite general in the presence of an optical lattice but we will focus on the free dispersion, {\it i.e.} $\eps_\bk=\bk^2/2m$). Here $n_0=\mu/g$ is the condensate density,  with $\mu$ the chemical potential and $mg$ is the dimensionless interaction strength. We use the convention $\hbar=k_B=1$ through the paper.
In the mean field approximation, interactions are only included between
non-condensed bosons and the condensate;
clearly, Bogoliubov level theory is inadequate as it does not include
dissipation. 

One can view this dissipation as arising from interactions between
thermal particles.  
Treating such interactions in a manner which leads to analytically
tractable dynamics
is not straightforward
\cite{Blakie2008}.
Thus, following the precedent of time dependent Ginsburg-Landau
theories, one introduces a ``noise" term 
\cite{Ullah1990}
via a bath \cite{Tan2004}.
The dynamics can be derived exactly and the calculations of the response
functions such as the structure
factors is then precise and fully consistent.
We note that other sources of dissipation which enter into the actual experiments
cannot be ruled out, as cold atoms systems are subject to lasers and other
probes and are not truly isolated.

Two obtain the appropriate analytic properties, one introduces 
two kinds of bosonic modes 
$\hat W^{(\dag)}_{j,\bk}$ and $\hat V^{(\dag)}_{j,\bk}$, with
Hamiltonian 
\begin{equation}
 \hat H_{{\rm bath}}= \sum_{j,\bk} \big[\omega_{j,\bk}\hat W^\dag_{j,\bk}\hat W_{j,\bk}+\nu_{j,\bk}\hat V^\dag_{j,\bk}\hat V_{j,\bk}\big].
\end{equation}
Here the index $j$ represents the bath degrees of freedom.
This bath then interacts with the system of interest via
\begin{equation}
\hat H_{\rm c}= \sum_{j,\bk} \big[\eta_{j,\bk}^* \hat W^\dag_{j,\bk} \hat \psi_\bk +\zeta_{
j,\bk} \hat V^\dag_{j,-\bk} \hat \psi^\dag_\bk+ h.c.\big],
\end{equation}
where $\eta_{j,\bk}$ and $\zeta_{j,\bk}$ represent generalized coupling constants.
The coupling 
is expected to
take one particle from the bath and put it into the system of
interest (or the opposite). Additional processes involve
a particle of the system and one of the bath falling into the condensate, and
the converse.

The equations of motion of the fields $i\partial_t\hat \psi_\bk(t)=\big[\psi_\bk,\hat H_f\big]$ and $i\partial_t\hat \psi^\dag_{-\bk}(t)=\big[\psi^\dag_{-\bk},\hat H_f\big]$ 
can be written
\begin{equation}
 \begin{split}
  i\partial_t\hat \psi_\bk(t)  =& \w_\bk \hat \psi_\bk(t)+ g_f n_0 \hat  \psi^\dag_{-\bk}(t)+\hat D_\bk(t)\\&-i \int_{t_0}^{t}ds\gamma_\bk(t-s)\hat\psi_\bk(s) ,\\
 i\partial_t\hat \psi_{-\bk}^\dag(t)  =& -\w_\bk\hat \psi_{-\bk}^\dag(t)- g_f n_0 \hat  \psi_{\bk}(t)-\hat D^\dag_{-\bk}(t)\\&-i \int_{t_0}^{t}ds\gamma_{-\bk}(s-t)\hat\psi^\dag_{-\bk}(s),
 \end{split}
\end{equation} 
where we have formally solved the equations of the bath operators. Here
before the quench,
the Hamiltonian with interaction strength $g_i$, in contact with
the bath is 
$\hat H_i = \hat H_{{\rm Bog}}(g_i)+\hat H_{{\rm bath}}
+\hat H_{\rm
c}$, while
after an instantaneous quench the Hamiltonian consists
of 
$\hat H_f = \hat H_{{\rm Bog}}(g_f)+\hat H_{{\rm bath}}+\hat H_{\rm c}$.
While it is not essential, as in previous work
\cite{Natu2012,Hung2012}
we neglect the time variation of the condensate density $n_0$ (which fixes the chemical potential $\mu_{i/f}=n_0 g_{i/f}$).
We define
$\w_\bk=\eps_\bk-\mu_f+2g_fn_0$, $\hat D_\bk(t)=\sum_j \eta_{j,\bk} e^{-i\w_{j,\bk}t}\hat W_{j,\bk}(0)+\sum_j \zeta_{j,\bk} e^{i\nu_{j,\bk}t}\hat V^\dag_{j,\bk}(0)$
 and $\gamma_\bk(t) =\int_\w \Sigma_2(\bk,\w) e^{-i\w t}$ with
$\int_\w=\int d\w/(2\pi)$. We define
$\Sigma_2(\bk,\w)=2\pi\sum_j \Big[|\eta_{j,\bk}|^2 \delta(\w-\w_{j,\bk})- |\zeta_{j,\bk}|^2
 \delta(\w+\nu_{j,\bk})\Big]$.

Here, $\hat D_\bk(t)$ plays the role of a random force operator and $\gamma_k(t)$ 
reflects the damping. The relaxation to equilibrium will be insured by the satisfaction of
the fluctuation-dissipation relation 
\begin{equation}
 \Big[\hat D_\bk(t),\hat D^\dag_\bk(s)\Big]=\gamma_\bk(t-s).
\label{eq_commut}
\end{equation}

In a more abstract form, our central result is a time dependent Bogoliubov equation
which now includes a damping term.
The equations of motion can be formally solved by introducing a matrix  $M_\bk(t)$ 
which will depend, via $\gamma_\bk(t)$,
 on the precise form of $\Sigma_2(\bk,\w)$, such that
\begin{equation}
\begin{split}
 \begin{pmatrix}
        \hat\psi_\bk(t)\\\hat\psi^\dag_{-\bk}(t)
       \end{pmatrix}
=& M_\bk(t)\begin{pmatrix}
        i\hat\psi_{\bk,0}\\ i\hat\psi^\dag_{-\bk,0}
       \end{pmatrix}
\\&+\int_0^{t}ds M_\bk (t-s)\begin{pmatrix}
        \hat D_\bk(s)\\ -\hat D^\dag_{-\bk}(s)
       \end{pmatrix}. \end{split} \label{eq_ev} \end{equation}


The 
simplest and most conventional
choice is to choose a so-called Ohmic bath (with spectral density of the bath,
$\Sigma_2(\bk,\w)$, proportional 
to $\w$ at small frequency) with the proper high-frequency regularization 
$\Sigma_2(\bk,\w)=2\Gamma_\bk \w/(1+\w^2/\Omega^2)$.  
Here $\Omega$ is a high-energy 
cut-off. Note that $\G_\bk$ is dimensionless. On physical grounds, we expect this damping parameter to increase with increasing interaction.

For this Ohmic bath, the time-evolution matrix $M_\bk(t)$ is given by (we dismiss a highly oscillating part 
with frequency of the order of $\Omega$ which does not play any role in the time 
evolution) \begin{equation}
 M_\bk(t)=e^{-\bg_\bk t}\begin{pmatrix}
        \frac{\cos(\bE_\bk t)-\frac{i\bw_\bk}{\bE_\bk}\sin(\bE_\bk t)}{(i-\Gamma_\bk)} & -\frac{\bgn}{\bE_\bk}\sin(\bE_\bk t) \\
	\frac{\bgn}{\bE_\bk}\sin(\bE_\bk t) &  \frac{\cos(\bE_\bk t)+\frac{i\bw_\bk}{\bE_\bk}\sin(\bE_\bk t)}{(i+\Gamma_\bk)}
       \end{pmatrix}, \label{eq_Matrix} \end{equation}
where we define the ``damped parameters" 
$\bw_\bk=\w_\bk/(1+\G_\bk^2)$, $\bgn=g_fn_0/(1+\G_\bk^2)$, $\bg_\bk=\G_\bk 
\bw_\bk$, $\bE_\bk^2=\big[E^2_{\bk,f}-\G_\bk^2 (g_fn_0)^2\big]/(1+\G_\bk^2)$, with $E^2_{\bk,\alpha}=\eps_\bk(\eps_\bk+2g_\alpha n_0)$ is the Bogoliubov energy.

 The matrix $M_\bk(t)$ can be recast as a sum of exponentials $\exp\Big[-(\bg_\bk\pm i\bE_\bk)t\Big]$ with complex frequencies. At large momenta, we find $(\bg_\bk\pm i\bE_\bk)\simeq (\G_\bk \eps_\bk\pm i E_\bk)/(1+\G_\bk^2)$ (assuming that $\G_\bk$ is bounded at large $|\bk|$, which is physically reasonable). The oscillations will therefore be typically given by the Bogoliubov energy, with expected damping at high-$|\bk|$ proportional to $\eps_\bk t$, even at short time. This is supported by experiment, as discussed below.
One can observe that, depending on the bath parameter $\G_\bk$, there may exist a 
range of momenta such that $\bE_\bk^2\leq 0$, implying that 
the dynamics is over-damped and $M_\bk(t)$ will not produce oscillations. 
This phenomenon, beyond Bogoliubov theory, may well have
been observed in \cite{Hung2012} for a quench to higher interaction strength 
at low-momenta.

\begin{figure*}
\includegraphics[width=2.0in,clip]
{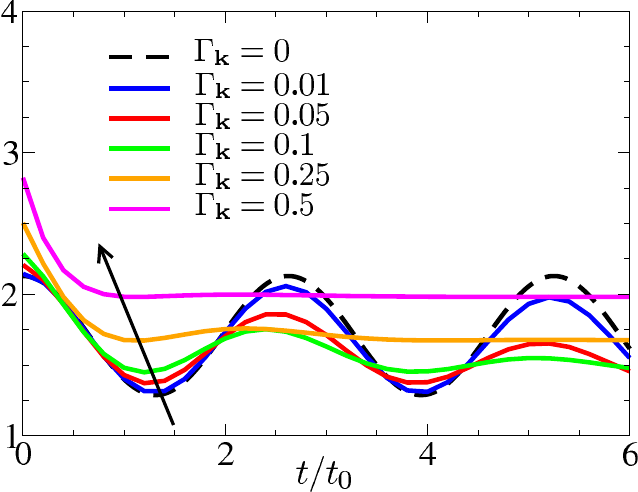}
\includegraphics[width=2.1in,clip]
{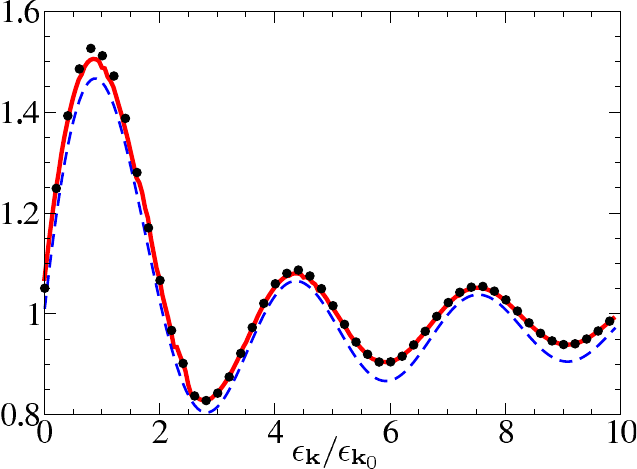}
\includegraphics[width=2.1in,clip]
{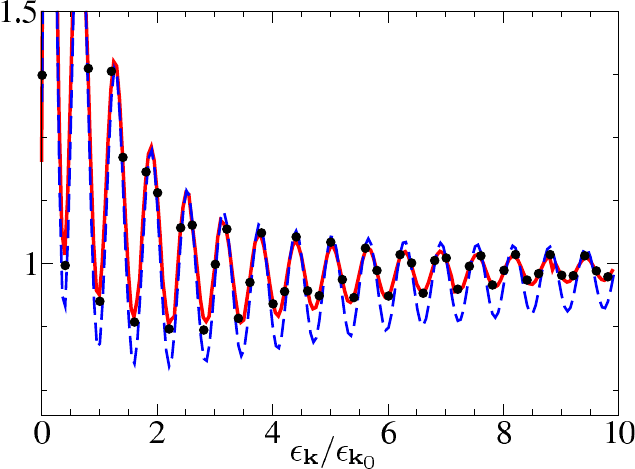}
\caption{(left) Structure factor at fixed time $t=2t_0$ (see text) as a function of momentum $k/k_0$, for different values of the damping parameter $\Gamma_{\bk}$, in the case of a quench up with $g_f/g_i\simeq1.8$. The arrow indicates the apparent up-shift of the frequency of the oscillations. Bogoliubov theory corresponds to $\Gamma_{\bk}=0$ (black dashed line). (center) Structure factor for quench down
at $t=t_0$ as a function of $\eps_\bk/\eps_{\bk_0}$ with $\Gamma_\bk=0.01$. The red curve is our result, the blue dashed curve Bogoliubov theory and the symbols a ``phenomenological fit'' (see text). (right) same as the central figure, but for $t=5t_0$.}
\label{fig:1}
\end{figure*}

\textit{Out-of-equilibrium correlation functions}
We now specify the out-of-equilibrium observables we will study.
Because our model Hamiltonian is quadratic, it is reasonably straightforward to
derive the four correlation functions 
involving combinations of 
$\psi$ and $\psi^\dag$.
Here we will concentrate on the density-density correlation functions, 
which are easily accessed in current cold atoms experiments \cite{Hung2011a}. The 
density operator is defined in momentum space as
$  \hat\rho_\q(t)=\sum_k \hat a^\dag_{\bk+\q}(t)\hat a_\bk(t)$, 
where 
$\hat a_\bk=\sqrt{n_0}\delta_{\bk,0}+\hat \psi_\bk$.
The different correlation 
functions will involve the usual time ordered normal 
($ G_\bk(t,s)$) 
and anomalous 
($F_\bk(t,s)$)
Green's functions.
Because of the quench, these
Green's functions
depend separately on their two time arguments.
We find as expected
that in the
long time limit, when the system reaches its new equilibrium state, they become
functions of only $t-s$. This is a non-trivial test of the current
theory which reflects the
energy dissipation mechanism.

%
%
The density-density correlation function $\chi_\q(t,s)=\mean{\hat
\rho_\q(t)\hat \rho_{-\q}(s)}-\mean{\hat \rho_\q(t)}\mean{\hat \rho_{-\q}(s)}$,
(with $t\geq s$) is related to the structure factor  
$S_\q(t)=\chi_\q(t,t)/n_0$.
It can similarly be written in terms
of the Green's functions so that
$ \chi_\q(t,s)=-n_0 \Big\{ G_\q(t,s)+G_\q(s,t)+F_\q(t,s)+F_\q^\dag(t,s)   \Big\}
+\sum_\bk \Big\{G_{\bk+\q}(s,t) G_\bk(t,s)+F^\dag_{\bk+\q}(t,s)F_\bk(t,s)\Big\}$,
where the sum is over all $\bk$ different from $0$ and $-\q$. Because the condensate is macroscopically occupied, we will neglect the second term in brackets in our numerical calculation of the structure factor \cite{Natu2012,Hung2012}. 


It is convenient to define the 4-vector $\Upsilon(t,s)=\Big\{F_\bk(t,s),G_\bk(s,t),
G_\bk(t,s),F^\dag_\bk(t,s)\Big\}$. 
These component Green's functions can be evaluated by solving 
equation \eqref{eq_ev}  
 which yields ($t>s$)
\begin{equation}
 \Upsilon(t,s)
=\int_0^t du\int_0^s dv \calM_\bk(v)\otimes\calM_\bk(u)\cdot \Upsilon_i(t-s-u+v),
\label{eq_correl}
\end{equation}
where the matrix $\calM_\bk(v)=\int_0^v M_\bk(v-v').M_{\bk,i}^{-1}(v') dv'$ 
and $M_{\bk,i}^{-1}(t)$ is the inverse of the $M_\bk$ with $g_f \to g_i$, with the definition  $\int_0^v M_\bk(v-v').M_{\bk}^{-1}(v')dv'=\delta(v)$. We have introduced the vector $\Upsilon_i(t)=\Big\{\calF_{\bk,i}(t),\calG_{\bk,i}(-t),\calG_{\bk,i}(t),\calF^\dag_{\bk,i}(t)\Big\}$, corresponding to the equilibrium Green's functions of the initial
Hamiltonian.
These can be defined using the 
the equilibrium
bosonic spectral functions
 \footnote{
The spectral function $\calA_{\bk,i}(\w)$ and its anomalous counterpart
$\calL_{\bk,i}(\w)$ are given in terms of the
Green's function
$G^{-1}_{\bk,i}(\w)= \w-\w_\bk-\Sigma_{\bk}(\w+i0^+)$ with the self-energy
$\Sigma_{\bk}(z)=\int_{\w'} \frac{\Sigma_2(\w')}{z-\w'}$, 
by
\begin{equation}
  \calA_{\bk,i}(\w)=
\frac{|G_{\bk,i}(\w)|^{-2}\Sigma_2(\bk,\w)-(g_i n_0)^2
\Sigma_2(\bk,-\w)}{|\calD_\bk(\w)|^2}
\end{equation}
\begin{equation}
\calL_{\bk,i}(\w)=
g_in_0\frac{\Sigma_2(\bk,\w)G^{*-1}_{\bk,i}(-\w)-G^{*-1}_{\bk,i}(\w)
\Sigma_2(\bk,-\w)}{|\calD_\bk(\w)|^2},
\end{equation} 
where $ \calD_\bk(\w)=G_{\bk,i}(-\w)G^*_{\bk,i}(\w)-(g_in_0)^2$. Note that the spectrum is
gapless, as $\calD_{\bk=0}(0)=0$ and that $\calA_{\bk,i}(\w)$ has the sign of $\w$
as required for bosons. In order to respect the commutation relation at equal time,
we have $\int_\w \calA_{\bk,i}(\w)=1$ and $\int_\w \calL_{\bk,i}(\w)=0$. }.
Our equations correspond to Bogoliubov theory in
the absence of a bath.

While Eq.~\ref{eq_correl} may seem formally complex, the physics it contains needs to
be emphasized.  Importantly, this equation is consistent with the fluctuation-dissipation 
theorem. Stated alternatively, it is consistent with the proper asymptotic
(long time) regime which requires that the final state of the system be time independent. It can be shown directly from 
Eq.~\ref{eq_correl} that the long time limit of 
$ \Upsilon(t+\tau,t)\to \Upsilon_f(\tau) $, which means that at long times the
proper equilibrium normal and anomalous Green's function of the quenched 
Hamiltonian $\hat H_f$ are obtained.
It should not be presumed that one can view this final state as a convolution of
a damping term and simple Bogoliubov theory; the asymptote of the structure factor
(for example) is finite, as a new equilibrium
phase is reached, so that the
oscillations are not simply damped out.

\textit{Numerical results} We discuss now the numerical solution of the previous equations. We introduce the characteristic momentum $k_0=\sqrt{n_0 m g_i}$ and time $t_0^{-1}=k_0^2/2m$. Note that $k_0$ is the inverse healing length of the condensate at $t=0$. 
The left panel of Figure \ref{fig:1} shows the evolution in time of the structure factor at a fixed $|\bk|$ after the quench-up for different values of the bath parameter $\G_\bk$ chosen to be independent of $|\bk|$ for simplicity. We also plot the results from Bogoliubov theory which corresponds to the case $\G_\bk=0$. 
Here we stress that because of damping
there is no single frequency observed for each $\mathbf{k}$.
Nevertheless, an
important effect of the bath (indicated by the arrow)
is the shift toward earlier time of the first extrema of the oscillations, leading to an apparent frequency increase.
This shift was observed in \cite{Hung2012} for a quench-up (where the effect of the bath is expected to be more important in the dynamics) and cannot be explained by Bogoliubov theory. This effect was not seen for a quench-down, which might be expected, as in this case $\Gamma_\bk$ is smaller.

 The central and right panels of Figure \ref{fig:1} 
plot $S_\bk(t)$ at fixed time $t=t_0$ and $t=10t_0$ for small $\Gamma=0.01$ in
our theory as compared with Bogoliubov theory.
For the latter, at long time one sees that
$S_\bk(t)$ oscillates faster and faster while never reaching a new equilibrium. 
We find a moderately successful phenomenological fit to our calculations
at low damping with 
$S^{ph}_\bk(t)=S_{\bk,f}+f_\bk(t)\Big(\big[1+(E^2_{\bk,i}-E^2_{\bk,f})/(E_{\bk,f})\sin(E_{\bk,f} t)^2\big]S_{\bk,i}-S_{\bk,f}\Big)$, 
where $S_{\bk,i}$ ($S_{\bk,f}$)  is the initial (final) equilibrium structure factor. 
For $|\bk|\gg g_f n_0$, we observe from our theory
that $f_\bk(t)=\exp(-(|\bk|/K(t))^2)$, where $K(t)=k_0\sqrt{\frac{(1+\Gamma_\bk^2) t_0}{\Gamma_\bk t}}$.
Fits to our theory are shown in the central and right panels
of Figure \ref{fig:1}. It should be stressed that for these small
values of the damping, as illustrated here, there is not yet a signature
of an apparent shift in frequency.

\begin{figure}[h!]
\includegraphics[width=2.0in,clip]
{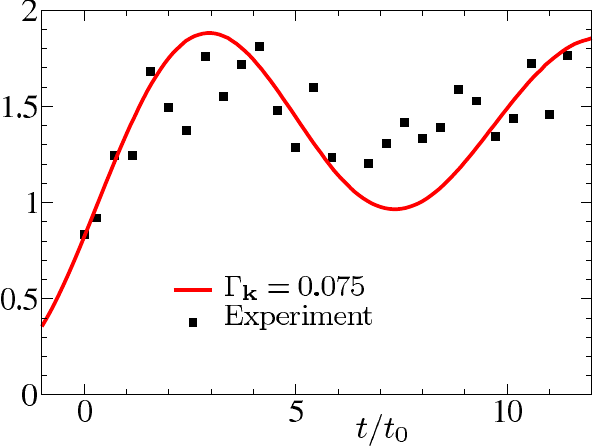}
\includegraphics[width=2.0in,clip]
{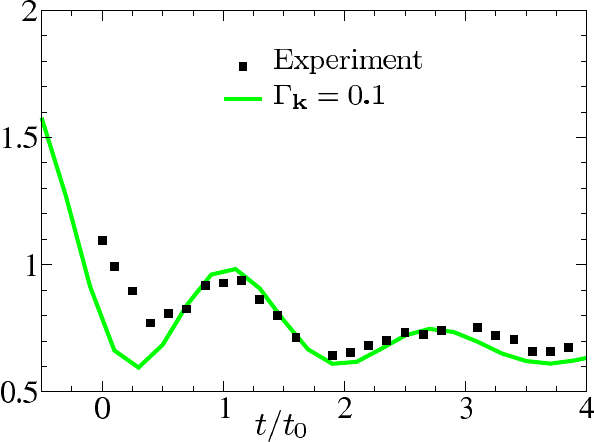}
\caption{Structure factor at fixed momentum as function of time. (left) quench down $g_f/g_i\simeq0.3$ and $k=0.9/\mu m$ with $k_0=1.6/\mu m$ and $t_0=0.7 ms$. (right) quench up $g_f/g_i\simeq2.4$ and $k=1.2/\mu m$ with $k_0=1/\mu m$ and $t_0=2 ms$.}
\label{fig:2}
\end{figure}

\textit{Experimental results} 
We have reported
\cite{Hung2012}
the experimental observation of
oscillatory behavior of the density structure factor associated with
a sudden quench in a 2D Bose system in the almost-pure- superfluid phase
of a cesium atomic gas. The emphasis of the present paper is on damping effects we
observe in
these oscillations. These are in contrast to an extensive literature
 focusing on shortcuts to adiabaticity after a fast change of the experimental parameters,
in particular of 1D gases, (see for instance \cite{Torrontegui2012}). 

The data points in Figures 2 represent
the measured equal-time structure factor as a function of time at
fixed $|\bk| =0.9\mu m^{-1}$ for
the final interaction strength $g_f = 0.19$ and for
a quench up with $g_f/g_i\simeq2.4$ (Fig. 2b)  and a quench down
(Fig. 2a) with $|\bk| =1.2\mu m^{-1}$ and $g_f/g_i\simeq0.3$.
Importantly, in contrast to other experimental
studies and deliberate quench protocols \cite{Schaff2011}, dissipation is evident
and
Sakharov oscillations will
eventually be damped out in the steady state.
We do not address this asymptotic long time
regime because our focus is on reasonably short times
where the trap geometry is not important.

We plot as solid lines in Figure \ref{fig:2}, 
theoretical curves with the same microscopic parameters. The color-coded
curves represent the regime of moderate damping for different values of
$\Gamma = 0.1$ (in a quench up) and $\Gamma = 0.075$ (in a quench down).
Since our focus is on semi-quantitative comparisons, we have allowed a global shift of the y-axis
as well as the x-axis (to take care of time uncertainties on the order of
$\lesssim 1ms$, due to time lapse between the quench and the detection).

In overall comparison between theory and experiment we find that
$\Gamma_{\mathbf k} \approx 0.1 $ describes the current data set. From this
observation, one cannot
yet characterize the microscopic nature and origin of dissipation, but our 
work should be viewed as a first step in the process.
Additional experiments and more systematic comparisons can be anticipated in
future. Nevertheless, with this approach, we can, however, address several features of
the experiments which were difficult to understand within strict
Bogoliubov theory such as the apparent frequency shifts and the damping both
in time and $\mathbf{k}$. 
A general theme of this paper has been to focus on damping effects.
Nevertheless, our out of equilibrium studies necessarily have implications
on the 
equilibrium
behavior (of the structure factor, say) due to the important
fluctuation-dissipation relation.
In summary, in this paper we have demonstrated that the introduction of
a Leggett-Caldeira bath for a Bose superfluid implies that
the dynamics can be derived exactly and the calculations of the response
functions such as the structure
factors is then precise and fully consistent.


\vskip3mm
A.R. thanks J. Bonart for discussions. This work is supported by NSF-MRSEC Grant
0820054.
C.L. and C.C. acknowledge support from NSF Grant No. PHY-0747907 and
under ARO Grant No. W911NF0710576 with funds from the DARPA OLE
Program.

\bibliography{/Users/adamrancon/Dropbox/Articles/bibli_bosons,/Users/adamrancon/Dropbox/Articles/bibli_RG,/Users/adamrancon/Dropbox/Articles/bibli_disorder,/Users/adamrancon/Dropbox/Articles/bibli_OutEq,/Users/adamrancon/Dropbox/Articles/bibli_BCSBEC,/Users/adamrancon/Dropbox/Articles/bibli_diverse}

\end{document}